\documentstyle[aps,twocolumn,epsf]{revtex}

\begin{document}
\title{ Monte Carlo Estimate of Finite Size Effects \\
in Quark-Gluon Plasma Formation}
\author{Andy~Gopie and Michael~C.~Ogilvie}
\address{ Department of Physics, Washington University, St. Louis, MO 63130}

\date{\today}

\maketitle

\begin{abstract}
Using lattice simulations of quenched QCD
we estimate the finite size effects
present when a gluon plasma equilibrates in a slab geometry,
{\it i.e.}, finite width but large transverse dimensions.
Significant differences are observed in the free energy density for the
slab when compared with bulk behavior. A small shift in the critical
temperature is also seen. The free energy required to liberate
heavy quarks relative to bulk is measured using
Polyakov loops; the additional free energy required is
on the order of $30-40 ~MeV$ at $2-3~T_c$.
\end{abstract}

\pacs{11.10.Wx 12.38.Gc 12.38.Mh}

\narrowtext

\section{INTRODUCTION}
\label{s1}

The formation of a quark-gluon plasma in a central heavy ion collision
is generally assumed to take place 
in a coin-shaped region roughly 1 fermi in width, with
radius comparable to the radii of the colliding nuclei, which is to
say several fermi.
While lattice gauge theory has given us information about bulk
thermodynamic behavior, finite size effects have up to now been
studied using simplified, phenomenological models.
We study via lattice gauge theory simulations
the behavior of a gluon plasma restricted to a slab geometry, with
the longitudinal width much smaller than the transverse directions.
This inner region is heated to temperatures above the bulk deconfinement
temperature, surrounded by an outer region which is kept at a temperature
below the deconfinement temperature,
providing confining boundary conditions for the
inner region.
Details of this work are given in \cite{Gopie}.

Measurements of the equilibrium surface tension $\alpha_0$
of pure SU(3) lattice gauge
theory (quenched lattice QCD) show that the dimensionless ratio
$\alpha_0/T_c^3$ is small.
For the case of an $N_s^3 \times 4 $ lattice,
$\alpha_0/T_c^3 \approx 0.0292(22)$.\cite{Iwasaki}
A simple estimate of surface tension effects on the transition
temperature
can be obtained from a simplified model in which only volume
and surface terms appear, as in the bag model.\cite{Cleymans}
Based on these simple considerations,
finite size effects due to the surface
tension should be small.

Other contributions to finite size effects come from a variety
of sources.
In the case of systems with non-abelian symmetries,
global color invariance produces an additional finite volume effect
which will not be considered here.
\cite{RedlichTurko}\cite{Turko}\cite{Skagerstam}
In general, finite size effects lead to a rounding of the transition.
\cite{BinderLandau}
This can be taken into account in the bag model 
by a Maxwell construction, leading to mixed phases
and a broadened critical region.
A recent treatment for the quark gluon plasma can be
found in
\cite{Spieles}.
We have attempted to avoid these finite volume effects by making the
transverse dimensions large.

\section{METHODOLOGY}
\label{s2}

In lattice calculations, finite temperature is introduced by the choice
of $N_t$, the extent of the lattice in the (Euclidean) temporal direction.
The relation of physical temperature $T$ to $N_t$ and the lattice
spacing $a$ is simply $T = 1 /{N_t a}$. The lattice spacing $a$ implicitly
depends on the gauge coupling $\beta = 6/g^2$ in a way determined by the
renormalization group equations.
To lowest order in perturbation theory, the relation is given by
\begin{eqnarray}
a \Lambda_L =  \left( {\beta \over {2 N b_0}} \right)^{ b_1/2b_0^2}
exp [- \beta / 4 N b_0 ] 
\end{eqnarray}
where $\Lambda_L$ is renormalization group invariant and
the renormalization group coefficients $b_0$ and $b_1$ are given by
\begin{eqnarray}
b_0 = { 11 N \over {48 \pi^2} }\ \ , \ \ 
b_1 = {34 \over 3 } \left( {N \over {16 \pi^2}} \right)^2
\end{eqnarray}
In analyzing our data, we used the 
renormalization group results given in reference \cite{Boyd},
which are determined directly from lattice simulations,
and contain non-perturbative information about the
renormalization group flow.

By allowing the coupling constant $\beta$ to
vary with spatial location, a spatially dependent temperature
can be introduced.
We have chosen the temperature interface to be sharp,
in such a way that the lattice is
divided into two spatial regions, one hotter and one colder.
The quenched approximation simplifies the role
of the cold region, because below $T_c$,
the dominant excitation at low energies is the scalar glueball.
Temperatures near $T_c$ are 
smaller than glueball masses by about a
factor of four, 
so glueballs play no essential role in the thermodynamics, and
the pressure in the hadronic phase is essentially zero.
We thus expect the slab thermodynamics to be largely
insensitive to the precise temperature of the region outside the
slab, as long as it is sufficiently low.
In full QCD, this insensitivity to the outer temperature
would not hold, due to pions.
Note that the role of boundary conditions here is quite
different from that in bubble nucleation. In that case, both
$\beta_{in}$ and $\beta_{out}$ are taken to be near $T_c$.
\cite{Kajantie}\cite{Huang2}

The slab is given a
fixed lattice width, $w=6a$, rather than of fixed physical width.
Since $N_t = 4$, $w T$ is fixed at $3/2$.
At higher temperatures, the width in physical units
is somewhat smaller than the
longitudinal size of the plasma formation region expected in heavy
ion collisions.
While the use of equilibrium statistical mechanics to study gluon plasma
properties during the early stages of plasma formation may appear
suspect, a simple estimate using the Bjorken model \cite{Bjorken}
shows that when a coin-shaped region of
width 1 fermi has expanded to 1.5 fermi, the variation in temperature
is only from $0.8 T_0$ at the center of the coin to $T_0$ at its edges.

The free energy density $f$ for the slab is obtained using the standard
method \cite{Engels} of integrating the lattice action with respect to
$\beta$. We use a convenient convention for the sign of $f$
that is opposite the usual one. In the bulk case, $f$ is then identical
to the pressure $p$.
\begin{eqnarray}
{f \over T^4}|_{\beta_{out}}^{\beta} =
N_t^4 \int_{\beta_{out}}^{\beta} d\beta'
\left[ \langle S \rangle_T - \langle S \rangle_0 \right]
\end{eqnarray}
where $ S  =  (1/N) Re Tr\ U_p$.
As in the bulk case, it is necessary to subtract the
zero-temperature expectation value from the finite temperature
expectation value, in this case
using the same pair of $\beta$ values.
In general, quantum field theories with boundaries develop divergences that
are not present in infinite volume or with periodic boundary conditions.
Such divergences would require additional boundary counterterms.
Symanzik \cite{Symanzik}
has shown to all orders in perturbation theory
that in the case of $\phi^4$ with so-called
Schrodinger functional boundary conditions that the theory is
finite in perturbation theory after adding all possible boundary
counterterms of dimension $d \leq 3$ consistent with the symmetries
of the theory.
It is generally believed that this result applies as well to all
renormalizable field theories and general boundary conditions,
but a proof is lacking. Luscher {\it et al.} \cite{Luscher}
have shown for gauge theories
that at one loop no new divergences are introduced
by Schrodinger functional boundary conditions. This is consistent
with the non-existence of gauge-invariant local fields of dimension
$ \leq 3$ in pure Yang-Mills theory.

In order to take advantage of the data on bulk thermodynamics
provided by the Bielefeld group \cite{Boyd}, we worked consistently
with lattices of overall size $16^3 \times 4$.
The values used for each subtraction come from
$16^4$ lattices with identical values of $\beta_{in}$ and
$\beta_{out}$.
The value of $\beta_{out}$ was held fixed at $5.6$ while
$\beta_{in}$ varied from $5.6$ to $6.3$.
For comparison, the bulk transition for $N_t = 4$ occurs at
$\beta_c (N_t=4, N_s=16) = 5.6908 (2)$
$\beta_c (N_t=4, N_s=\infty) = 5.6925 (5)$.
\cite{Boyd}

\section{FREE ENERGY OF GLUONS}
\label{s3}

Figure 1 shows the free energy density $f/T^4$ versus $T/T_c$ compared with
the bulk pressure.
The free energy in the slab is lower than the bulk value by almost
a factor of two at $2 T_c$. It appears that the slab value is
slowly approaching the bulk value, but other behaviors are also
possible. 
Calculations of the finite-temperature contribution to the
Casimir effect for a free Bose field contained between two plates
show that $f/T^4$ has a non-trivial dependence on the dimensionless
combination $w T$.\cite{Mehra} \cite{Plunien}
It is natural to ask if the corrections to the free energy seen here
can be accounted for by the conventional Casimir effect.
A straightforward calculation of the free energy of a
non-interacting gluon gas confined to a slab shows an increase in
the free energy density over the bulk value by a factor of about 1.63
at $wT = 3/2$. The Casimir effect cannot explain the
reduction of the free energy observed in our simulations.

\begin{figure}[htb]
\epsfxsize=75mm \epsfbox{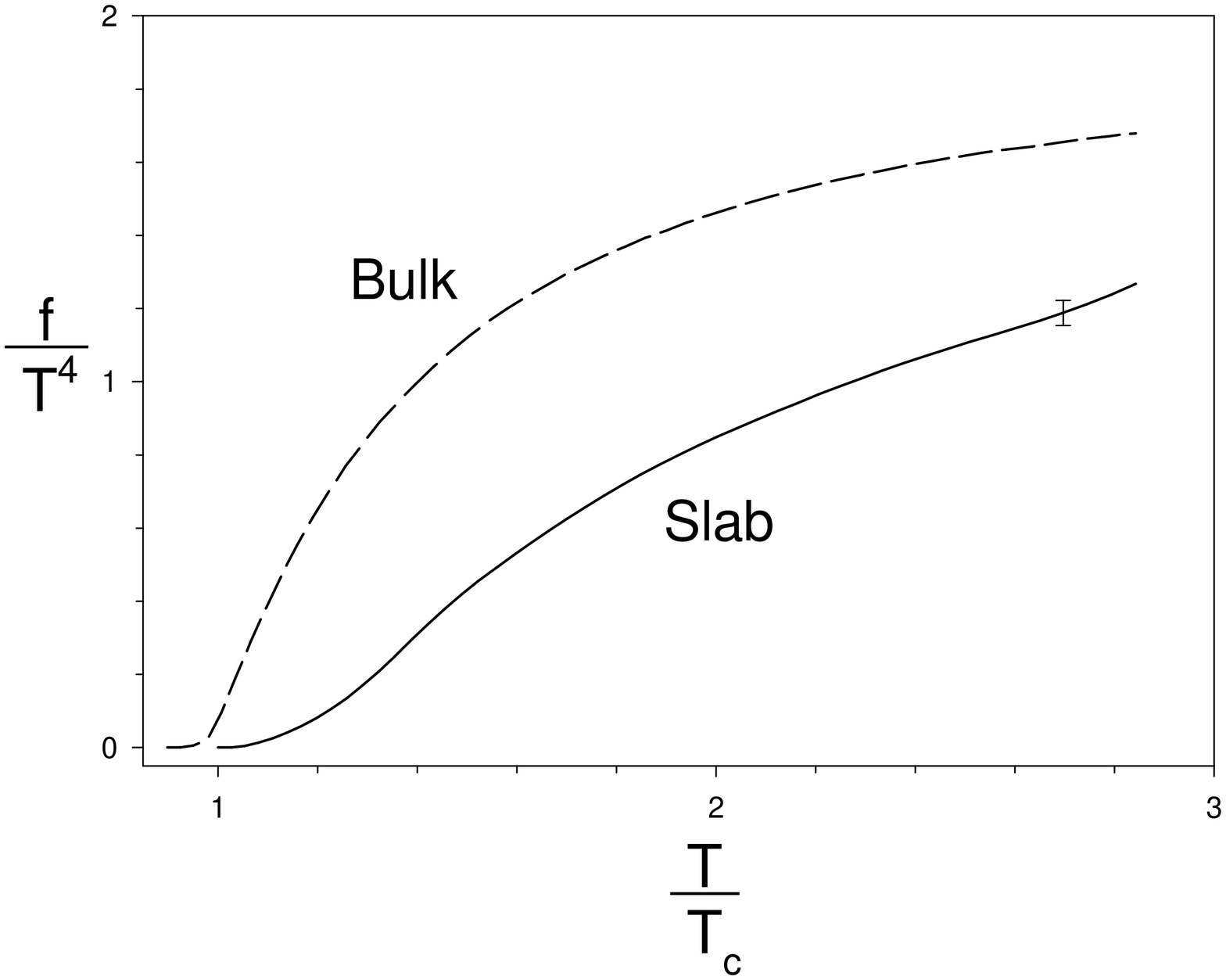}
\caption{Slab free energy compared with bulk pressure.}
\label{fig:fig1}
\end{figure}

A consistency check was performed on the surface effects.\cite{Huang} 
The free energy density was calculated for a system at $\beta_{in} = 6.0$
by performing simulations with
$\beta_{in}$ fixed at $6.0$ and $\beta_{out}$ varying from $5.6$ to $6.0$.
Combining these results with the bulk data of reference \cite{Boyd}
creates a path equivalent to varying $\beta_{in}$ while holding
$\beta_{out}$ fixed.
For $\beta_{in} = 6.0$ and $\beta_{out}=5.6$, 
this gives $f/T^4 = 0.65 \pm 0.04$,
to be compared with $f/T^4 = 0.69 \pm 0.03$
for the direct calculation.
The major source of systematic error lies with the choice of
boundary conditions for the slab, here set by $\beta_{out}$.
We have estimated the effects of varying $\beta_{out}$ by
performing simulations at $\beta_{in} = 6.2$
and $\beta_{out} = 5.5$ on $16^3 \times 4$ and $16^4$ lattices.
These results suggest that lowering $\beta_{out}$
from $5.6$ to $5.5$ reduces the free energy by roughly
10 percent at $\beta_{in} = 6.2$.

\section{SURFACE TENSION}
\label{s4}

We define an effective surface tension $\alpha(w,T)$ by
\begin{eqnarray}
f = p - 2 \alpha(w,T) / w
\end{eqnarray}
where the notation $\alpha(w,T)$ recognizes that the surface tension
$\alpha$ does depend on the width of the slab and the internal
and external temperatures.
The factor of $2$ occurs because the slab has two faces.
In the limits where $T$ approaches $T_c$ and $w$ goes to infinity,
this quantity approaches $\alpha_0$.
Figure 2 shows $\alpha(w,T)/T^3$ versus $T/T_c$ for $w T = 3/2$;
representative error bars are shown.
\begin{figure}[htb]
\epsfxsize=75mm \epsfbox{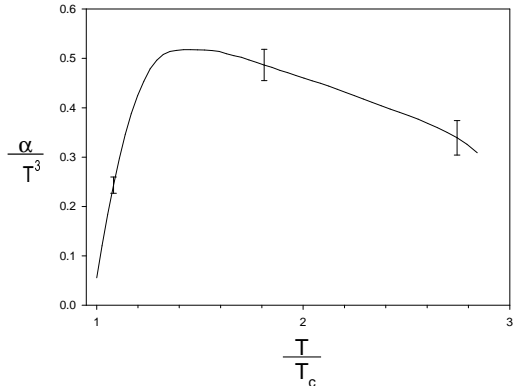}
\vspace{0.1in}
\caption{Non-equilibrium surface tension versus $T$.}
\label{fig:fig2}
\end{figure}
The value at $\beta=5.7$, $0.056 \pm 0.002$ is 
higher than the
value $\alpha_0 = 0.0292 \pm 0.0022$ given in reference \cite{Iwasaki}.
We attribute this to two effects:
in our case $\beta_{out}$ is
fixed at 5.6, whereas for equilibrium measurements it is
extrapolated to $\beta_c$, and our finite value of the width $w$ also
acts to increase $\alpha(w,T)$ over the equilibrium value as measured
in simulations at large $w$.
Away from the bulk critical point, $\alpha / T^3$ rises quickly
to a peak at about $1.4 T_c$, and then falls slowly as $T$ increases.
A large non-equilibrium surface tension has also been observed
in measurements of the equilibrium surface tension, where these
effects were obstacles to obtaining $\alpha_0$.\cite{Huang}

\section{FREE ENERGY OF QUARKS}
\label{s5}

The Polyakov loop defined by
\begin{eqnarray}
P(\vec{x}) =  (1/N_c) Tr~ {\cal P} \exp
\left[  i \int^{1/T}_0 A_0(\vec{x}, \tau)~ d\tau \right]
\end{eqnarray}
is the order parameter for the deconfinement transition in pure (quenched)
gauge theories. In the case of $SU(N)$, there is a global
$Z(N)$ symmetry which ensures at low temperature that the
expectation value $\langle Tr P \rangle$ is $0$. At sufficiently
high temperatures, this symmetry is spontaneously broken.
The expectation value of the Polyakov loop can also be
interpreted in terms of the free energy of an isolated,
infinitely heavy quark $F_Q$:
\begin{eqnarray}
\left< P(\vec{x}) \right> =  exp \left[ -F_Q(\vec{x})/T \right]
\end{eqnarray}
In the low-temperature confined phase, $F_Q$ is taken to be
infinite, whereas in the high temperature phase it is finite.
Direct extraction of $F_Q$ from computer simulations is problematic,
because the expectation value has a multiplicative, $\beta$-dependent
ultraviolet divergence.
This divergence can be eliminated when comparing
bulk expectation values to those in finite geometries.
We define
\begin{eqnarray}
\Delta F_Q(\vec{x}) =  -T~ ln \left[ P_{slab}(\vec{x})/P_{bulk} \right]
\end{eqnarray}
as the excess free energy required to liberate a heavy quark
in the slab geometry relative to bulk quark matter at the
same temperature. 
This technique can also be used in, {\it e.g.},
a spherical geometry, which is relevant for nucleation.
\cite{Ogilvie}\cite{Kajantie}

In figure 3, we show the 
expectation value for the Polyakov loop versus
z measured in lattice units for several values of $\beta$.
Each curve is normalized by dividing the values of the Polyakov
loop by the bulk expectation value at the corresponding value of
$\beta$.
Error bars are shown only for even values of z.
\begin{figure}[htb]
\epsfxsize=75mm \epsfbox{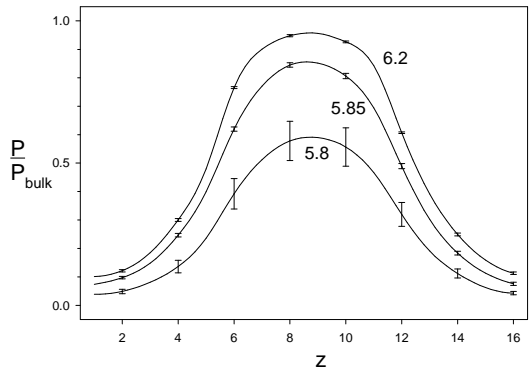}
\caption{Polyakov loop versus $z$.}
\label{fig:fig3}
\end{figure}
It is clear that a significant change occurs between 
$\beta = 5.8 ~~~(T= 1.23~T_c)$ and $\beta = 5.85 ~~~(T= 1.36~T_c)$.
For larger values of $\beta$, $\Delta F_Q$ diminishes to a
value  of approximately $30 - 40 ~~MeV$ in the middle of the slab.
In table 1, we list $\beta$, $T$, slab width in fermis, width
of the core in fermis, and $\Delta F_Q$ in MeV for representative values.
The width of the core is calculated by interpolating the Polyakov loop
profiles and determining the region where the slab expectation value
is greater than 80\% of the bulk value.
All conversions
to physical units are performed by 
taking the string tension $\sigma$ to be $(425 ~ MeV )^2$,
which implies $T_c = 254~MeV$.\cite{Boyd}

\begin{table}
\caption{$\beta$, $T$, slab width in fermis, width
of the core in fermis, and $\Delta F_Q$ in MeV for representative values.}
\begin{tabular}{lllll}
$\beta$ & $T$ (MeV) & w (fm.) & $w_{core}$ (fm.) & $\Delta F_Q$ (MeV) \\
\tableline
5.8 & 313 & 0.95 & 0 & 169 \\
5.85 & 346 & 0.86 & 0.40 & 57 \\
6.0 & 455 & 0.65 & 0.48 & 41 \\
6.2 & 624 & 0.47 & 0.40 & 31 \\
\end{tabular}
\end{table}

\section{CONCLUSIONS}
\label{s6}
There are significant deviations in the
slab geometry from bulk behavior and the ideal gas law,
arising from a strong non-equilibrium
surface-tension. This non-equilibrium surface tension can be
an order of magnitude greater
than the equilibrium value.
Surface tension effects also produce a
mild elevation of the apparent critical temperature.
Measurement of Polyakov loop expectation values relative to bulk
shows that the suppression of heavy quark production
due to the slab geometry is small.

There are good reasons to call our result an estimate rather than
a calculation. Although lattice gauge theory simulations of bulk
behavior can be made arbitrarily accurate in principle, in this case
there is some uncertainty in the precise exterior boundary conditions
appropriate, and indeed in the applicability of equilibrium
thermodynamics at this early stage of quark-gluon plasma formation.
However, this seems like the best estimate available now,
and further refinements are possible.

We have not yet explored the nature of the phase transition, which
will require some care.
One interesting possibility is that the order of the transition might
change as the width changes. The deconfinement transition in bulk quenched
finite temperature QCD is in the universality class of the three-dimensional
three-state Potts model, which has a first-order phase transition.
As the width of the slab becomes commensurate with the correlation length
near $T_c$, the phase transition should
cross over to the universality class of the
two-dimensional three-state Potts model.
The two-dimensional three-state Potts model has a second-order phase
transition, so it is possible that the order of the transition may
change.\cite{ShnidmanDomany}
The correlation length at the bulk
transition is known to be large \cite{Fukugita},
so it is likely that
the transverse correlation length in the gluonic sector
is much larger in the
slab geometry than in bulk, even if crossover does not
take place.

The system studied here has some interesting features amenable
to theoretical analysis. In the outer region, the Polyakov loop
will decay away from the interface as $ exp [ - \sigma r / T_{out} ] $,
where $\sigma$ is the string tension.
In the inner region, the Debye screening length sets the scale
for the Polyakov loop. We are currently working on 
a theoretical model of this system based on perturbation theory
which has these features.

\section*{ACKNOWLEDGEMENTS}
We wish to thank the U.S. Department of Energy for financial support under
grant number DE-FG02-91-ER40628.

\end{document}